\pdfoutput=1
\documentclass[twocolumn,
               showpacs,
               preprintnumbers,
               nofootinbib,
               prd,
               superscriptaddress,
               10pt,
               notitlepage,
               aps]{revtex4-2}
               
\usepackage{graphicx,amssymb,amsmath,amsthm,amsfonts,epsfig,mathtools}

\usepackage[utf8]{inputenc}
\usepackage{graphicx}
\usepackage{dcolumn}
\usepackage{bm}
\usepackage{amsmath}
\usepackage{color}
\usepackage[dvipsnames]{xcolor}
\usepackage{hyperref}
\hypersetup{colorlinks=true, citecolor=MidnightBlue,linkcolor=CornflowerBlue, urlcolor=CornflowerBlue, linktocpage=true}
\usepackage{xfrac}
\usepackage{siunitx}
\usepackage{soul}
\newcommand{\beqn}{\begin{eqnarray}}
\newcommand{\eeqn}{\end{eqnarray}}
\newcommand{\beq}{\begin{equation}}
\newcommand{\eeq}{\end{equation}}

\def\gbar{\bar{g}}
\def\gnn{\bar{g}_{nn}}
\def\gnn{\bar{g}_{nn}}

\begin{document}

\title{Pervasiveness of the breakdown of self-interacting vector field theories}

\begin{abstract}
Various groups recently argued that self-interacting vector field theories lack a well-defined time evolution when the field grows to large amplitudes, which has drastic consequences for models in gravity and high energy theory. Such field amplitudes can be a result of an external driving mechanism, or occur intrinsically, due to large values of the field and its derivatives in the initial data. This brings a natural question: is small amplitude initial data guaranteed to evolve indefinitely in these theories in the absence of an outside driving term? We answer this question in the negative, demonstrating that arbitrarily low amplitude initial data can still lead to the breakdown of the theory. Namely, ingoing spherically symmetric wave packets in more than one spatial dimensions grow as an inverse power of the radius, and their amplitudes can generically reach high enough values where time evolution ceases to exist. This simple example further establishes the pervasiveness of the  pathology of self-interacting vector field theories.
\end{abstract}

\author{Andrew Coates}
\email{acoates@ku.edu.tr}
\affiliation{Department of Physics, Ko\c{c} University, \\
Rumelifeneri Yolu, 34450 Sariyer, Istanbul, Turkey}

\author{Fethi M. Ramazano\u{g}lu}
\email{framazanoglu@ku.edu.tr}
\affiliation{Department of Physics, Ko\c{c} University, \\
Rumelifeneri Yolu, 34450 Sariyer, Istanbul, Turkey}

\date{\today}
\maketitle

\section{Introduction}
Self-interacting vector fields are not as well known as the massless vectors of electromagnetism and their massive version, the Proca field, but they have been used in many areas of physics. These include gravity and cosmology for dark energy and boson star models~\cite{Esposito-Farese:2009wbc,DeFelice:2016yws,DeFelice:2016cri,Heisenberg:2017hwb,Kase:2017egk,Ramazanoglu:2017xbl,Annulli:2019fzq,Barton:2021wfj,Minamitsuji:2018kof, Herdeiro:2020jzx,Herdeiro:2021lwl,Garcia-Saenz:2021uyv,Silva:2021jya,Demirboga:2021nrc,Doneva:2022ewd}, plasma physics for effective modelling of photon-photon and photon-plasma interactions~\cite{RevModPhys.78.591}, astrophysical superradiance~\cite{Conlon:2017hhi,Fukuda:2019ewf,dEnterria:2013zqi, Burgess:2020tbq} and beyond~\cite{Heisenberg:1936nmg,ATLAS:2017fur}. They have also been theoretically investigated in some detail, and all ghost-free generalizations of the Proca theory have been categorized under the name of \emph{generalized Proca theories}~\cite{Proca:1936fbw,Heisenberg:2014rta, Heisenberg:2016eld, Kimura:2016rzw, Allys:2015sht}.

Despite the existing body of work, in recent years it has been shown that, even the simplest nonlinear extensions of the Proca theory, can break down in finite time~\cite{Esposito-Farese:2009wbc,Clough:2022ygm,Mou:2022hqb,Coates:2022qia,Coates:2022nif}. This occurs when the differential equations governing the dynamics lose their hyperbolic nature at large field amplitudes, and the evolution cannot continue any further. The problem arises from the fact that the dynamics of the vector field is governed by an \emph{effective metric}, not the spacetime metric, which is a function of both the spacetime metric and the vector field itself. The effective metric can become singular at finite values of the norm of the vector field, and such values can be achieved starting from healthy initial data for which the effective metric is Lorentzian~\cite{Coates:2022qia}.\footnote{There have been conflicting reports on where exactly the theory breaks down, see \textcite{Coates:2022nif} for a discussion.} Even though the breakdown is most intensively studied for a relatively simple extension of the Proca theory, the revealed pathology is likely to exist in any self-interacting vector field, hence the problem most likely occurs in all the specific theories we mentioned above. This simple case is also the effective field theory obtained by truncating the Abelian Higgs theory at the leading order, which provides further motivation for its study. 

Since the breakdown of time evolution occurs at relatively large amplitudes, existence of a mechanism that incites the growth of the field is an essential part of the well-posedness study of self-interacting vectors. Such growth can easily be achieved by an external driving source, one concrete example being the exponential growth of the vector due to superradiance near rotating black holes~\cite{Clough:2022ygm,Mou:2022hqb}. It is also possible to reach the breakdown point without any external coupling to the vector if one starts from small field amplitudes but with large initial time derivatives~\cite{Coates:2022qia}. In both cases, the breakdown of time evolution has been numerically observed, but it is also known that without an external driving source, the fields can evolve indefinitely for certain low-amplitude initial data, that is, the breakdown does not occur at all. 

An immediate question is whether self-interacting vectors can generically evolve indefinitely, without time evolution problems, if initial data is low amplitude both in field values and their derivatives. If this is the case, then one might argue that such theories can still be utilized freely to explain nature as long as the energy scale relevant for the time evolution breakdown is large enough, even without any recourse to viewing the theory as an effective one. Here, we demonstrate that this is \emph{not} the case. There are initial data configurations with arbitrarily small values for both the vector field amplitude and energy, whose evolution leads to breakdown even without any external coupling.

The said initial data is a spherical wave packet in more than one spatial dimension. Unlike plane waves, the amplitude of spherically symmetric wave packets change as inverse powers of the radius, for example they scale as $r^{-1}$ in $3+1$ dimensions. Thus, they can grow by arbitrarily large factors if they are ingoing, i.e. moving to smaller radii. This growth is well-known for massless vector fields, e.g. electromagnetism, and we will show that the nonlinear interactions do not alter the fundamental picture, which means the growth can eventually lead to the breakdown of time evolution as well. The appearance of this problematic behavior without any large initial amplitudes or external driving also demonstrates the pervasiveness of the pathology of self interacting vectors, and indicates that they are largely ruled out as fundamental theories. Nevertheless, we will also briefly discuss the situation if one interprets these theories as effective ones.

We use the ``mostly plus" metric signature $(-,+,\dots, +)$, with spacetime indices in Greek $\mu,\nu=0,1, \dots ,d$ and purely spatial indices in Latin $i,j=1,\dots , d$. $c=1$.

\section{Breakdown of Time Evolution}
Our summary of the breakdown of self-interacting vector field theories will follow \textcite{Coates:2022qia, Coates:2022nif}, however similar earlier results can also be found in \textcite{Esposito-Farese:2009wbc}. We investigate the action
\begin{align}\label{eq:action}
    {\cal L} = -\frac{1}{4} F_{\mu\nu}F^{\mu\nu} - \overbrace{\left( \frac{\mu^2}{2} X^2 + \frac{\lambda \mu^2}{4} \left( X^2\right)^2 \right)}^{V(X^2)} \ ,
\end{align}
where $F_{\mu\nu} = \nabla_\mu X_\nu - \nabla_\nu X_\mu$ and $X^2 =X_\mu X^\mu$ for the real vector field $X_\mu$. $\lambda=0$ is the Proca theory, hence the nonlinearity is controlled by this parameter. The spacetime metric $g_{\mu\nu}$ is nondynamical, and we will use it to lower and raise tensor indices, and define the connection. Even though our results will be specific for this action, they can be qualitatively generalized to other $V(X^2)$, and hold in the generic case.

The equation of motion arising from the action~\eqref{eq:action} is
\begin{align}\label{eq:eom}
    \nabla_\mu F^{\mu\nu} = \mu^2 z X^{\nu} \ ,
\end{align}
where $z=2V'/\mu^2=1+\lambda X^2$ and $V'=(dV/dX^2)$. This also implies the (generalized) Lorenz condition
\begin{align}\label{eq:lorenz}
    \nabla_\nu \nabla_\mu F^{\mu\nu} = 0\ 
    \Rightarrow\ \nabla_\mu \left(z X^\mu \right) = 0 
\end{align}
due to the antisymmetry of $F_{\mu\nu}$. Combining the two preceding equations, it is possible to write the principal part, the highest derivative terms, of the vector field equations as a wave operator acting on the vector~\cite{Clough:2022ygm,Coates:2022qia}\footnote{Strictly speaking, this is only possible in $1+1$D, however $\gbar_{\alpha\beta}$ still determines where the breakdown occurs in any dimension~\cite{Coates:2022qia}.}
\begin{align}\label{eq:gbar_eom}
    \gbar_{\alpha\beta}\nabla^\alpha \nabla^\beta X_\nu + \dots 
    = 0 \ , 
\end{align}
where lower order derivatives are  not explicitly shown. The striking feature of this last equation is that the wave operator is not the one that corresponds to the spacetime metric $g_{\mu\nu}$, but rather to a new \emph{effective metric}
\begin{align}
    \gbar_{\mu\nu} &= z g_{\mu\nu} + 2z' X_\mu X_\nu\ . \label{eq:g_eff} 
\end{align}

If $\gbar$ becomes singular or ceases to be Lorentzian, the principle part of the field equation no longer represents hyperbolic time evolution, which is sometimes called \emph{loss of hyperbolicity}~\cite{East:2022ppo,Corman:2022xqg,R:2022hlf}. This unwelcome prospect indeed occurs for finite values of $X_\mu$ when
\begin{equation}\label{eq:detg}
    \gbar = g \left(1+\lambda X^2 \right) \left(1+3\lambda X^2 \right) = g\ z\ z_3 = 0
\end{equation}
where $g = \det(g_{\mu\nu})$, $z_3 = 1+3\lambda X^2$. In general, the determinant of a metric can vanish due to coordinate effects, such as for the flat metric in spherical polar coordinates. In this case, however, it is explicitly known  that $\gbar=0$ is a curvature singularity~\cite{Coates:2022qia}. Hence, the effective metric becomes singular at $z=0$ or $z_3=0$, but it is trivial to show that the latter always occurs first when starting from small amplitude initial data. In summary, the breakdown of time evolution occurs when the vector field norm satisfies $X^2 = -1/(3\lambda)$.

\section{Spherically Symmetric Wave Packets}
A prerequisite for the breakdown of time evolution is the growth of $-\lambda X^2$ as we discussed, but how can this occur if the field is not driven by an external factor as in superradiance~\cite{Clough:2022ygm}, or the initial conditions provide a large momentum~\cite{Coates:2022qia}?\footnote{There is a third alternative, an intrinsic tachyonic instability which occurs for $\mu^2<0$. We only consider $\mu^2>0$ in this study.} A very simple answer is provided by spherical waves in $3+1$ dimensions.

Let us start with the linear Proca theory, $\lambda=0$, in flat spacetime in spherical polar coordinates
\begin{align}\label{eq:flat_metric}
    ds^2 = -dt^2 + dr^2 + r^2 dS^2\ ,
\end{align}
to understand the appeal of this physical example, where $dS^2$ is the standard metric on the two-sphere. Furthermore, consider a vector field which itself is spherically symmetric, and can be parametrized as
\begin{align}\label{eq:X_param}
    X_\mu=(X_t,X_r,0,0 ) = r^{-1} (\tilde{X}_t(t,r),\tilde{X}_r(t,r),0,0 )\ .
\end{align}

This leads to the field equations
\begin{align}\label{eq:spherical_eom}
    0 &= \left(-\partial_t^2 + \partial_r^2\right) \tilde{X}_t -\mu^2 \tilde{X}_t \\
    0 &= \left(-\partial_t^2 + \partial_r^2\right) \tilde{X}_r -\left(\mu^2+\sfrac{2}{r^2}\right) \tilde{X}_r\ , \label{eq:spherical_eom_r}
\end{align}
which are consistent with spherical symmetry, that is, if the angular components of the vector field and their time derivatives are zero at the initial time, they remain so forever.

$\tilde{X}_t$ obeys the Klein-Gordon equation, hence its normal modes simply behave as massive plane waves,
\begin{align}\label{eq:massive_wave}
    \tilde{X}_{t}(t,r) = \sum_k a_k e^{\pm ikr} e^{i\sqrt{k^2+\mu^2}t}\ .
\end{align}
The key element in this result is the $r^{-1}$ factor in Eq.~\eqref{eq:X_param} which increases the overall amplitude of the normal modes near the origin since $X_t = r^{-1} \tilde{X}_{t}$. The field equation~\eqref{eq:spherical_eom_r} for $\tilde{X}_r$ is slightly more complicated, but the Lorenz condition
\begin{align}
    r\partial_t \tilde{X}_t(t,r) &= \partial_r \left[ r \tilde{X}_r(t,r) \right]
\end{align}
guarantees that its modes have the same $r^{-1}$ factor.

The behaviour of the modes imply that a well localized, i.e., narrow enough, ingoing wave packet can grow by any given factor $N \gg 1$ if it moves from an initial position $r_0$ to $r_0/N$, which is also the well-known behavior of massless electromagnetic waves~\cite{Jackson99}. The narrowness is important for two reasons. First, note that the $r^{-1}$ factor does not imply that the field values diverge at $r=0$. The modes always superpose to lead to finite values, hence a wave packet does not blow up at the origin. Rather, the modes that superpose to form the wave packet behave in such a way that they cancel out at and near the origin. Hence, given the width of the wave packet $\Delta r$, the $r^{-1}$ behavior is valid as long as the final position is larger than the wavepacket width,
\begin{align}
    r_0/N \gtrsim \Delta r \ .
\end{align}

The second reason that makes the narrowness of the wave packet essential is dispersion. The dispersion relation for the massive wave of Eq.~\eqref{eq:massive_wave} is $\omega(k) = \sqrt{k^2+\mu^2}$. When $\omega/k$ is not a constant, different modes have different phase velocities $v_p(k)=\omega(k)/k$. As a result, any wave packet that is a superposition of such modes spreads out and its amplitude decreases as it evolves. Thus, very strong dispersion, in principle, can suppress the growth arising from the $r^{-1}$ factor present in individual modes. The effect of dispersion is weaker for $k \gg \mu$ where the massive dispersion relation mimics the dispersionless massless case of electromagnetism. For a wave packet of size $\Delta r$, the typical wave numbers in the (generalized) Fourier transform satisfy $k \sim 1/\Delta r$. This immediately suggests that dispersion is weak for \
\begin{align}
    \Delta r \lesssim \mu^{-1} \ ,
\end{align}
and we expect to see the $r^{-1}$ growth to dominate over the decreasing effect of dispersion on the amplitude.

Beyond arbitrarily low amplitudes, we can also show that arbitrarily low-energy configurations are sufficient for breakdown. Again consider a $3+1$ dimensional spherical wave of initial amplitude $1/N$ and width $\Delta r$ centred around the radius $r_0$, which loses hyperbolicity when it grows $\sim N$-fold in amplitude while going in. Since the volume of the packet is $\sim 4 \pi r_0^2 \Delta r$, we would want to minimize the initial radius to minimize the total energy. Recall that the solution is regular at $r=0$, hence we cannot choose $r_0$ to be arbitrarily small, and the amplitude increases only before the packet is reflected from the origin. Thus, the closest we can get to the origin while the amplitude is still growing is $\sim \Delta r$, meaning, the smallest initial radius we can start is $\sim N \Delta r$. In summary, a wavepacket of amplitude $N^{-1}$ where the energy is confined to a narrow region $\Delta r \to 0$ can initially be positioned around $r_0 \sim N \Delta r$. Naively, the energy density is dominated by the spatial gradient of $X_t$ whose contribution is $(N^{-1}/\Delta r)^2$, which ultimately means the total energy behaves as $\sim 4\pi (N \Delta r)^2 \Delta r\ (N^{-1}/\Delta r)^2 \sim \Delta r$. However, the constrained nature of the dynamics means the derivative terms cannot be freely specified, and are related to the field amplitudes in a nontrivial manner. Nevertheless, a more detailed calculation shows that the effect of this leads to even lower energies, see Appendix~\ref{sec:3p1} and~\ref{sec:energy}. In other words, we can use as little energy as we want by choosing a small enough $\Delta r$, completing our argument.

If the essential behavior we discussed for the linear theory ($\lambda=0$) is preserved for $\lambda \neq 0$, we expect a small ingoing spherical wave packet in $3+1$ dimensions to grow in amplitude and break time evolution when $X^2=-1/(3\lambda)$ is achieved. The $\lambda=0$ case only provides supporting evidence, albeit strong, since, nonlinear effects can potentially change the behavior of the field, especially near the breakdown. A second concern is that our Lorentzian metric~\eqref{eq:flat_metric} means $X^2 = -X_t^2+X_r^2$, hence $X^2$, which solely determines the breakdown, can be small due to a cancellation even if both $X_t$ and $X_r$ are growing as $r^{-1}$. In the following section, numerical computation will show that these concerns do not materialize, and ingoing spherical wave packets of self-interacting vectors indeed reach a point of loss of hyperbolicity.

\section{Numerical Results}
Since there is no known exact analytical solution for the nonlinear self-interacting case, we need to compute the time evolutions using numerical methods in order to confirm our arguments in the preceding section. We followed the methods of \textcite{Coates:2022qia} for the time evolution, which are in turn based on the $3+1$ decomposition of \textcite{Clough:2022ygm} and the general numerical relativity literature~\cite{gourgoulhon20123+1,Zilhao:2015tya}. Just like the linear case, initially spherically symmetric self-interacting vectors stay so during their time evolution so that the exact nonlinear dynamics effectively forms a $1+1$ dimensional system for $X_{t,r}$. See Appendix~\ref{sec:3p1} for further details.

We used initial data that is a spherically symmetric narrow Gaussian wave packet of $X_r$ located far from the origin, with vanishing $X_t$ and vanishing time derivatives for both field components. Quantitatively, this means $\Delta r \lesssim \mu^{-1}$ and $r_0 \gg \Delta r$, $\Delta r$ being the standard deviation and $r_0$ the position of the maximum of the Gaussian. This time-symmetric data means the wave packet splits into an ingoing and outgoing part during evolution, but our investigation concentrates solely on the growing ingoing piece.

\begin{figure}
\begin{center}
\includegraphics[width=.48\textwidth]{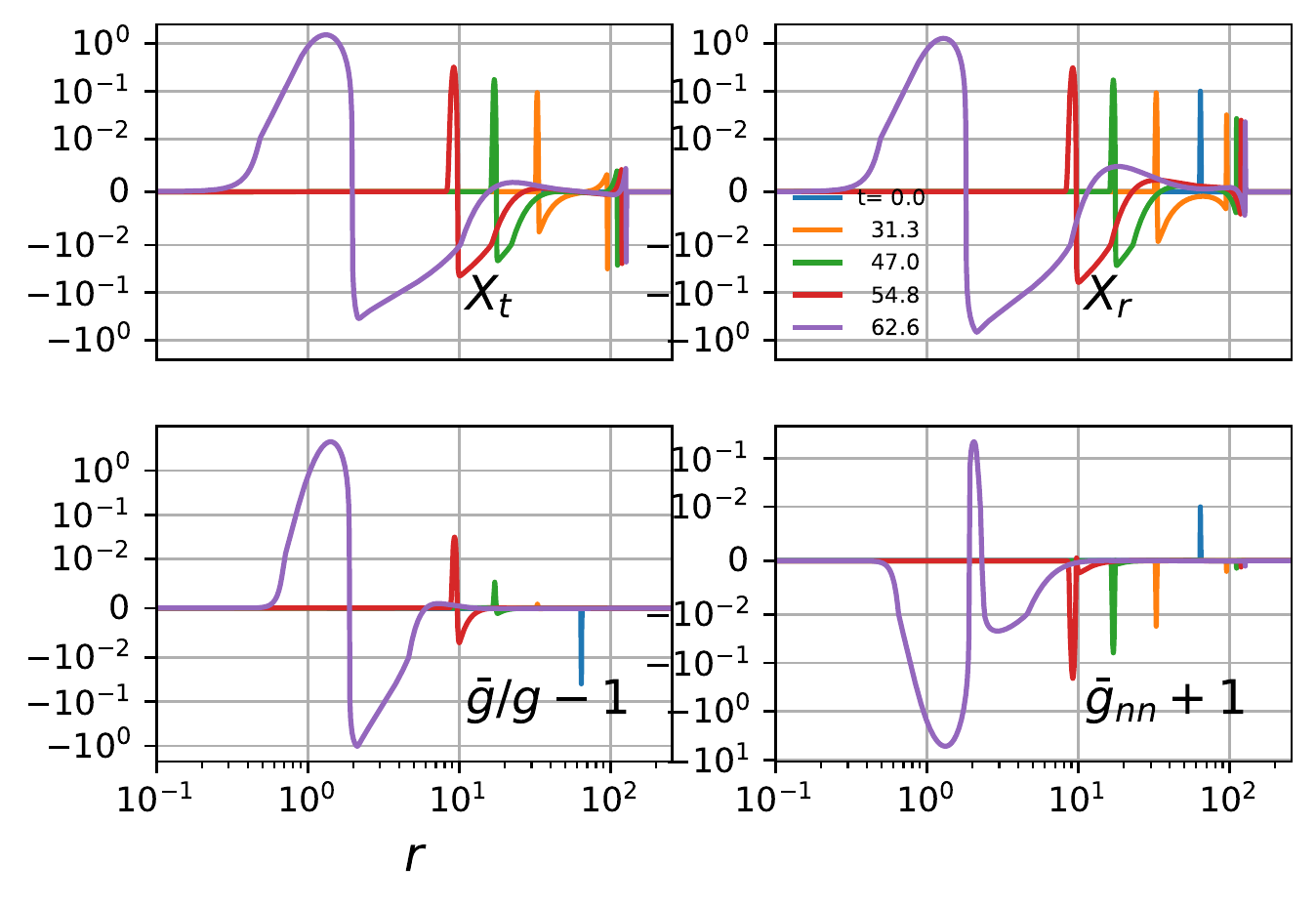}
\end{center}
\caption{Snapshots of $X_\mu$ and $\gbar_{\mu\nu}$ as a wave packet moves to smaller radii in flat $3+1$ dimensional spacetime, $\mu=0.1, \lambda=-1$. The initially-Gaussian wave packet starts with a small width, but it widens in late times as it moves inward (to the left) due to dispersion. The growth of the amplitude dominates nevertheless, leading to the growth of $\lambda X^2$ and the breakdown of time evolution, $\gbar=0$. This occurs for arbitrarily low initial amplitudes, provided the initial position of the pulse is far enough from the origin. The seeming discontinuity of the first derivative of the pulse at the amplitude of $10^{-2}$ is an artefact of the ``symlog'' scaling on the vertical axis, which is linear near zero and logarithmic elsewhere. $\gnn$ is the metric component in the normal-normal direction, see Eq.~\eqref{eq:eom1p1_long}.
}
\label{fig:3p1}
\end{figure}
A sample evolution that ends in the breakdown of time evolution is in Fig.~\ref{fig:3p1} for $\mu^2=0.1, \lambda=-1$. Overall, our expectations are confirmed. Namely, $X_{t,r}$ indeed grow as $r^{-1}$. Hence, our analytical arguments hold to a sufficient degree despite nonlinear effects. Moreover, and crucially, there is no significant cancellation in $X^2 = -X_t^2+X_r^2$ either, which grows as $r^{-2}$, and eventually causes the breakdown when $z_3 =1+3\lambda X^2=0 \Rightarrow \gbar=0$ is encountered. The vector field develops a sharp feature at this point, and the numerical computation crashes soon after.

Even though Fig.~\ref{fig:3p1}  is a single example, the behavior is robust in that any small amplitude, narrow wave packet grows as it moves in, up to the point it reaches the origin. Hence, by starting from a large enough distance, arbitrarily small initial amplitudes can lead to the breakdown of time evolution.

Another point to emphasize is that our results inform us about the whole $(\mu^2,\lambda)$ parameter space, despite using a single point on it.  One can always set $\mu^2$ to any positive value by scaling the spacetime coordinates, and $\lambda$ to any negative values by scaling the vector field amplitudes for $\lambda<0$. Hence, different parameter values would only change the absolute measurement units, but the fact that ingoing spherical wave packets can grow arbitrarily large remains the same in all cases. The case of $\lambda>0$ can be similarly understood by studying $\lambda=1$, which we believe to have the same behavior. However, the time evolution in this case necessarily encounters coordinate singularities for the numerical methods we utilize~\cite{Coates:2022qia}, hence we could not check this explicitly. 

Our analytical results readily generalize to $d+1$ dimensions for $d>1$, where ingoing spherically symmetric wave packets grow as $r^{-(d-1)/2}$ in field components $X_{t,r}$ and $r^{-(d-1)}$ in $X^2$. We did not check such cases numerically, but strongly suspect that the breakdown occurs for any $d$. It seems that $1+1$ dimensions is the most resilient to breakdown, where there is no growth for an ingoing wave packet. Hence, the pathalogies of self-interacting vectors become more apparent in higher dimensions.

\section{Discussion}
We explicitly demonstrated that the time evolution of self-interacting vectors can break down even when the initial data has arbitrarily small amplitudes and even in the simplest case of flat spacetime. Not relying on horizons, any specific energy input mechanism or initial data with large momentum demonstrates the problems with self-interacting vectors in a much simpler sense than before.

The quantum mechanical uncertainty principle and the collapse of an exceedingly narrow wave packet when gravitational interactions are turned on would be important in general. Nevertheless, it is striking that the theory of action~\eqref{eq:action}, when taken at face value classically, breaks down starting from such tiny perturbations. In some respects, the singularity of self-interacting vectors is similar to a general relativistic spacetime curvature singularity arising from a collapsing spherical shell, but the analogy is far from exact. We emphasize that gravity does not play any role in the breakdown in our example, the spacetime metric is flat and non-dynamical. Moreover, where time evolution ceases to exist is strictly controlled by the amplitude of the vector fields through $X^2$, and energy does not play any direct role, unlike gravitational collapse.

Even though we have shown how easily time evolution breaks down for self-interacting vectors, further study is needed to fully understand the nonlinear effects. No system is exactly spherically symmetric in nature, and the nonlinear effects of angular perturbations can only be seen in $3+1$ dimensional solutions without symmetry assumptions. Hence, understanding the effects of asymmetry will require more advanced numerical methods.

Finally, our analysis took action~\eqref{eq:action} at face value as a fundamental field. However, in some cases it is possible to view it as an effective field theory where the more complete theory can be pathology-free. In this perspective, the problematic theory can still be useful as long as its limitations are recognized~\cite{Aoki:2022woy,Barausse:2022rvg}. Investigating this approach will be one of the main topics of interest in the coming years.

\acknowledgements
We thank Will East for many valuable discussions and stimulating questions. F.M.R acknowledges support from T\"UB\.ITAK Project No. 122F097.

\appendix

\section{Equations for numerical time evolution}
\label{sec:3p1}
In order to obtain a time evolution scheme, we decompose the metric and the vector field into their spatial and time parts under the $3+1$ decomposition as
\begin{align}\label{eq:adm}
    ds^2 &= -\alpha^2 dt^2 +\gamma_{ij} (dx^i + \beta^i dt)(dx^j + \beta^j dt) \\
    X_\mu &= \phi\ n_\mu  + A_\mu \  ,  \ \phi = -n_\mu X^\mu \  ,  \ A_i=\left(\delta^{\mu}{}_{i} +n^\mu n_i \right)X_\mu \ . \nonumber
\end{align}
where the normal vector $n^\mu = \alpha^{-1}(1,-\beta^i)$ defines how spatial surfaces are located inside the spacetime, the \emph{foliation}. Further defining
\begin{equation}
    E_i = \left(\delta^{\mu}{}_{i} +n^\mu n_i \right)n^\nu F_{\mu\nu}\ 
\end{equation}
leads to the field equations~\cite{Clough:2022ygm,Coates:2022qia,Zilhao:2015tya}
\begin{equation}\label{eq:eom1p1_long}
  \begin{aligned}
    {\rm d}_t \phi &= -A^i D_i \alpha - \frac{\alpha}{\gnn} z \left(K\phi -D_i A^i \right) +\frac{\alpha}{\gnn} Z \\
     +\frac{2\lambda \alpha}{\gnn} &\left[A^i A^j D_i A_j -\phi \left(E_i A^i - K_{ij} A^i A^j + 2A^i D_i \phi \right) \right] \\
    {\rm d}_t A_i &=-\phi D_i \alpha - \alpha \left( E_i +D_i \phi \right)\\    
    {\rm d}_t E_i &= D^j \left[ \alpha \left(D_i A_j -D_j A_i \right) \right] \\ 
    &\phantom{=} +\alpha \left(KE_i -2K_{ij}E^j +D_i Z \right) 
    +\mu^2 z \alpha A_i
     \\ 
    {\rm d}_t Z &= -\alpha \left(\kappa  Z - {\cal C}\right) \\
    0&= D_i E^i + \mu^2 z \phi = {\cal C} \\
    z &= 1+\lambda A_i A^i -\lambda \phi^2\\
    \gnn &= n^\mu n^\nu \gbar_{\mu\nu} = -z+2 \lambda \phi^2 = -z_3 +2\lambda A_iA^i\ .
  \end{aligned}
\end{equation}
Note that ${\cal C}=0$, called the \emph{constraint equation} does not represent dynamical evolution, rather, it provides a necessary condition that has to be satisfied on all spatial surfaces, including the initial data surface. We also introduced the \emph{constraint damping} term $Z$, which ensures that the constraint ${\cal C}$ does not artificially grow due to numerical reasons~\cite{Zilhao:2015tya}. See standard sources~\cite{gourgoulhon20123+1} for the details of the lapse $\alpha$, shift $\beta$, induced metric $\gamma_{ij}$, extrinsic curvature $K_{ij}=-{\rm d}_t \gamma_{ij}/2\alpha$ and its trace $K=\gamma^{ij} K_{ij}$. Spatial (Latin) indices are raised and lowered with $\gamma_{ij}$, and the ``total derivative'' is ${\rm d}_t = \partial_t - {\cal L}_\beta$, ${\cal L}_\beta$ being the Lie derivative. 

The equations above imply that on a spherically symmetric static spacetime background, if the angular components of the vector field and their time derivatives vanish for initial data, than they vanish at all times, i.e. an initially spherically symmetric configuration always remains so. This means the dynamics is restricted to the time-radius sector, and is effectively $1+1$ dimensional.

In our computations, we consider the flat metric of Eq.~\eqref{eq:flat_metric} so that $\alpha=1$, $\beta^i=(0,0,0)$, $\gamma_{ij}={\rm diag} (1,r^2, r^2\sin^2\theta)$.

\begin{figure}
\begin{center}
\includegraphics[width=.48\textwidth]{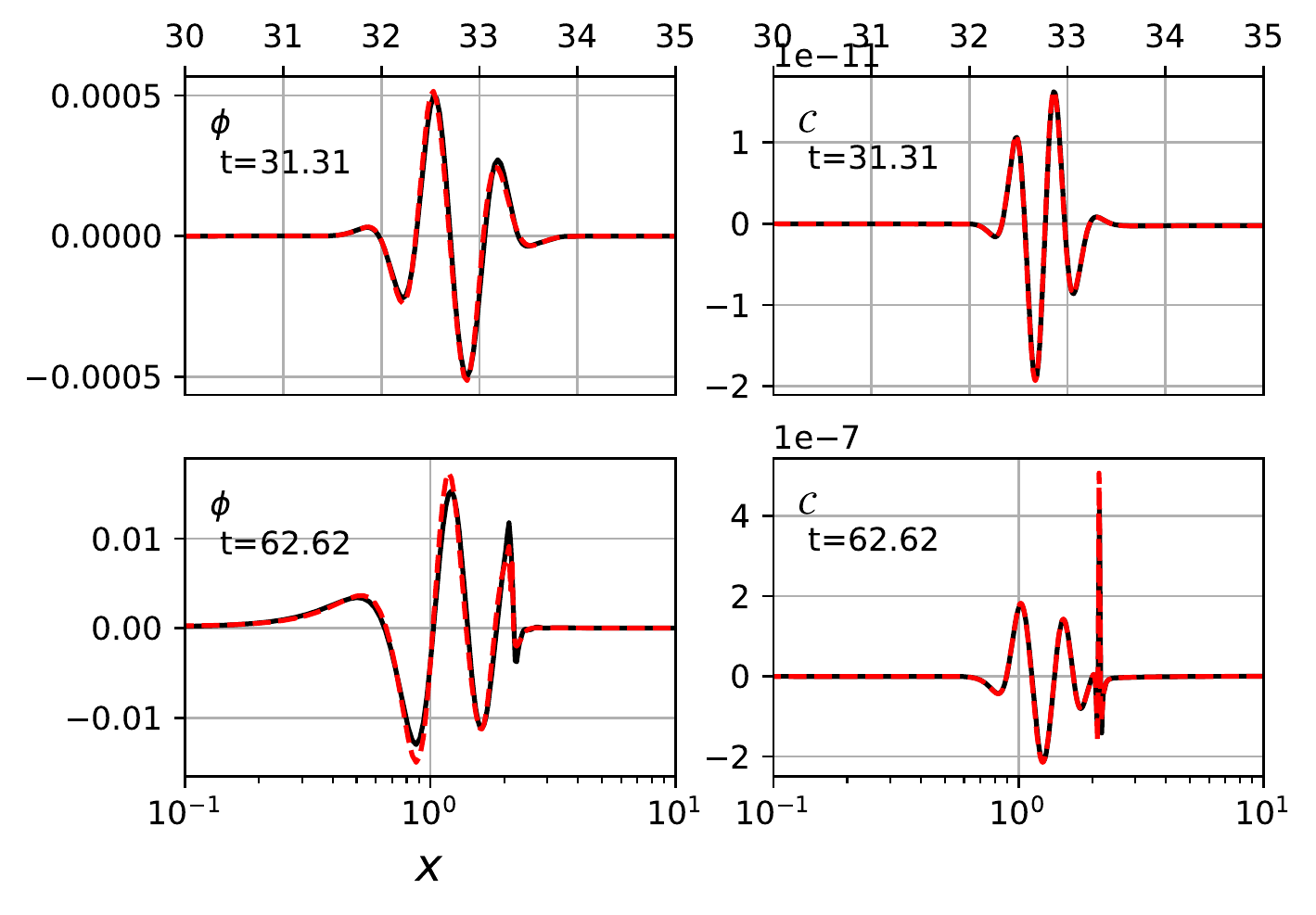}
\end{center}
\caption{Convergence of the numerical field values $f_{\Delta r}$ demonstrated by repeating the computation for step sizes $\Delta r = 2^{-6}$, $2\Delta r$ and $4 \Delta r$. 4\textsuperscript{th} order pointwise convergence is achieved in all cases. Left column: Comparison of the truncation error estimates $\phi_{4\Delta r}- \phi_{2\Delta r}$ (black) and $16(\phi_{2\Delta r}- \phi_{\Delta r})$ (red) at mid-time (upper) and end of computation (lower). Right column: Likewise comparison of the numerically computed constraints ${\cal C}_{2\Delta r}$ (black) and $16{\cal C}_{\Delta r}$ (red). Unlike $\phi$, the constraint itself converges to zero.
}
\label{fig:convergence}
\end{figure}
Our numerical setup is very similar to that of \textcite{Coates:2022qia}. We use the method of lines with $4^{th}$ order spatial derivatives and $4^{th}$ order Runge-Kutta integration in time. We used $6^{th}$ order Kreiss-Oliger dissipation~\cite{Kreiss73} with $\epsilon= 2^6\sigma_{\rm KO} =0.3$~\cite{shibata2015numerical} to avoid high frequency noise. For the sample evolution in Fig.~\ref{fig:3p1}, we used the spatial domain $r \in [\Delta r ,\ 256]$, spatial step size $\Delta r=2^{-6}$, Courant–Friedrichs–Lewy factor $\Delta t/\Delta r=0.25$, and constraint damping parameter $\kappa=1$. The initial wave packet is a Gaussian of the form
\begin{equation}
    X_r(t=0,r) = 0.1 \exp\left[- \frac{(r-64-2^{-8})^2}{2\ (0.25)^2} \right]
\end{equation}
with all other vector components and their derivatives vanishing. 

This initial data leads to values of exactly zero away from the center of the wave packet due to finite numerical precision. As a result of this and thanks to the finite propagation speed of the wave packet, the fields remain zero throughout the computation near $r=0$ and the outer boundaries. Under ideal circumstances, this means the boundary conditions imposed at the two ends do not affect the simulation. However, to avoid any spurious numerical effects moving in from the origin, we imposed the physical boundary conditions $\partial_r \phi =0 =\partial_r Z,\ A_r = 0 = E_r$ at $r=0$. The outer boundary does not come into causal contact with the ingoing wave, and we simply imposed the (unphysical) Neumann boundary conditions $\partial_r \phi =\partial_r Z=\partial_r A_r = \partial_r E_r=0$ there.


We repeat the computation with two and four times coarser grids, and observe 4\textsuperscript{th} order convergence as expected from our setup. Details can be seen in Fig.~\ref{fig:convergence}.

\section{Energy of the wave packet}
\label{sec:energy}
The stress-energy tensor derived from Eq.~\eqref{eq:action} is,
\begin{align}
    T^\alpha{}_\beta&=F^{\alpha\gamma} F_{\beta\gamma} - \frac{1}{4}\delta^{\alpha}{}_{\beta}F^{\gamma\delta}F_{\gamma\delta}\\
    &+\mu^2\left[X^\alpha X_\beta \left(1+\lambda X^2 \right) -\frac{1}{4}\delta^{\alpha}{}_{\beta} X^2 \left(2+\lambda X^2 \right) \right]  \nonumber 
\end{align}

For $X_{t} = C_{t} N^{-1} e^{(r-N\Delta r)^2/2(\Delta r)^2}$, the largest contribution to the energy in the $\Delta r \to 0$ limit \emph{naively} comes from the $\partial_r X_t$ terms, as
\begin{align}\label{eq:energy_total}
&\int_0^\infty - 4\pi r^2 T^0{}_0\mathrm{d}r= C_t^2 \pi^{3/2} \Delta r + \mathcal{O}\left(\Delta r^3\right)  
\end{align}
which is linear in $\Delta r$ as we discussed. However, the constraint equation, ${\cal C}=0$, implies that it is not sufficient for $X_{t,r}$ to be localized in a small region for the energy to be also localized in the same region. For example, if $X_r=0$ for the given $X_t$, than the only nonvanishing derivative term  $F_{rt}=E_r$ does not necessarily vanish at infinity (due to the implied value of $\partial_t X_r$ by ${\cal C}=0$). This means the energy is not confined, and we do not have a narrow wave packet to begin with. It turns out, however, when the energy is indeed confined, the dominant term can be much lower than Eq.~\eqref{eq:energy_total}. For example, if $X_{r} = C_{r} N^{-1} e^{(r-N\Delta r)^2/2(\Delta r)^2}$ and $X_t=0$ as in Fig.~\ref{fig:3p1}, then $E_r=0$, and the dominant term in the total energy becomes $2 C_r^2  \pi^{3/2} \mu^2 \Delta r^3$.


\bibliography{references_all}

\end{document}